\documentclass[a4paper]{VisionStyle}
\usepackage{epsfig}

\begin{document}

\title{X-ray powerful diagnostics for highly-ionized plasmas: He-like ions}

\author{D. Porquet\inst{1,5} \and R.\,Mewe\inst{2} 
\and  J.S. Kaastra\inst{2} \and J. Dubau\inst{3,5} \and  A.J.J. Raassen\inst{4,2}} 

\institute{
Service d'Astrophysique, CEA Saclay, 91191 Gif-sur-Yvette Cedex, France
\and 
SRON, Sorbonnelaan 2, 3584 CA Utrecht, The Netherlands
\and 
LSAI, U.M.R. 8624, CNRS, Universit{\'e} de Paris Sud, 91405 Orsay Cedex, France
\and 
Astronomical Institute ``Anton Pannekoek"
, 1098 SJ Amsterdam, The Netherlands 
\and 
LUTH, F.R.E. 2462, Observatoire de Paris, section de Meudon, 92195 Meudon Cedex
}
\maketitle 

\begin{abstract}

The calculations of the ratios of the Helium-like
 ion X-ray lines
from \ion{C}{v} to \ion{Si}{xiii}  are revisited in order to apply the results
to density, temperature and ionization process diagnostics of data from
high-resolution spectroscopy of the new generation of X-ray satellites:
 {\sl Chandra} and {\sl XMM-Newton}.
 Comparing to earlier computations, Porquet \& Dubau (\cite{dporquet-WB2:Porquet00}), 
the best experimental values are used for 
radiative transition probabilities. 
The influence of an external radiation field (photo-excitation),
 the contribution from unresolved dielectronic satellite lines 
and the optical depth are taken into account.
 These diagnostics could be applied to collision-dominated plasmas 
(e.g., stellar coronae), photo-ionized plasmas 
(e.g., ``Warm Absorber'' in AGNs), and transient plasmas (e.g., SNRs).

\end{abstract}

\section{Introduction}
  
With the advent of a new generation of X-ray satellites (Chandra and XMM-Newton),
 X-ray spectroscopy for \linebreak extra-solars objects with unprecedented spectral resolution 
and high S/N is now possible for the first time. 
Various plasma diagnostics are accessible such as those based on the line ratios 
of He-like ions. 
 The wavelength ranges of the RGS (6-35 {\AA}), of the LETGS (2-175 {\AA}), and 
of the HETGS (MEG range: 2.5-31 {\AA}; HEG range: 1.2-15 {\AA}) contain the Helium-like 
line "triplets'' from \ion{C}{v} (or \ion{N}{vi} for the RGS, and for the HETGS-HEG) 
to \ion{Si}{xiii} (Table~\ref{table:lambda}). 
The ratios of these lines was already widely used for solar plasma diagnostics 
(e.g., Mewe \& Schrijver \cite{dporquet-WB2:Mewe78a}, \cite{dporquet-WB2:Mewe78b}, 
\cite{dporquet-WB2:Mewe78c}; 
Doyle \cite{dporquet-WB2:Doyle80}; Pradhan \& Shull \cite{dporquet-WB2:Pradhan81}).\\

The analysis of the He-like ``triplet'' is a powerful tool in the analysis of the 
high-resolution spectra of a variety of plasmas such as: \\
$\bullet$ collisional plasmas: {\it e.g., stellar coronae 
(OB stars, late type stars, active stars, ...)}\\
$\bullet$ photo-ionized or hybrid plasmas (photo-ionization + collisional ionization): 
{\it e.g., ``Warm Absorber'' (in AGNs), X-ray binaries, ...}\\
$\bullet$ out of equilibrium plasmas: {\it e.g., SNRs, ...}

\section{Diagnostics}

\begin{figure}
\epsfig{file=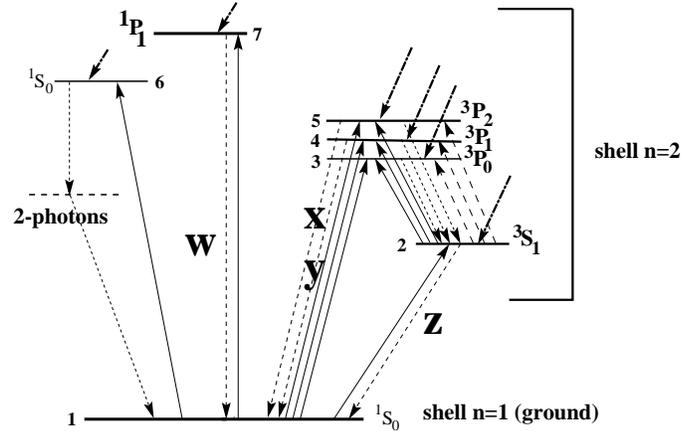,width=8.9cm}
\caption{Simplified level scheme for Helium-like ions. 
{\bf w} (or r), {\bf x,y} (or i), and {\bf z} (or f): resonance, intercombination, and forbidden lines, 
respectively. {\it Full upward arrows}: collisional excitation transitions, 
{\it broken arrows}: radiative transitions (including photo-excitation from 
2\,$^{3}$S$_{1}$ to 2\,$^{3}$P$_{0,1,2}$ levels, and 2-photon continuum from  
2\,$^{1}$S$_{0}$ to the ground level), and  {\it thick skew arrows}:  
recombination (radiative and dielectronic) plus cascade processes.}
\label{fig:gotrian}
\end{figure}

\indent In the X-ray range, the three most intense lines of Helium-like ions (``triplet'') are: the 
{\it resonance} line ($w$, also called $r$: 1s$^{2}$\,$^{1}S_{\mathrm{0}}$ -- 1s2p\,$^{1}P_{\mathrm{1}}$), 
the {\it intercombination} lines ($x+y$, also called $i$: 1s$^{2}$\,$^{1}S_{0}$ -- 1s2p\,$^{3}P_{2,1}$) and 
the {\it forbidden} line ($z$, also called $f$: 1s$^{2}$\,$^{1}{S}_{0}$ -- 1s2s\,$^{3}{S}_{1}$). 
They correspond to transitions between the $n$=2 shell and the $n$=1 ground-state shell 
(see Figure~\ref{fig:gotrian}).\\ 

Gabriel \& Jordan (\cite{dporquet-WB2:Gabriel69}) introduced the 
techniques to determine, from the ratios $R$ and $G$, 
electron density and temperature of the Solar corona:
\begin{equation}\label{eq:RG}
R~(n_e)~=~\frac{z}{(x+y)}~~~~~~~~~~~~~~G~(T_e)=\frac{(x+y)+z}{w}
\end{equation}

\begin{table}[!h]
\caption{Wavelengths in \AA\, of the three main X-ray lines of \ion{C}{v}, \ion{N}{vi}, \ion{O}{vii}, 
\ion{Ne}{ix}, \ion{Mg}{xi} and \ion{Si}{xiii} (from Vainshtein \& Safronova 1978).}

\begin{center}
\begin{tabular}{c@{\ }c@{\ }c@{\ }c@{\ }c@{\ }c@{\ }c@{\ }c@{\ }}
\hline
\hline
{\small line}                 &    {\small label}   & {\small \ion{C}{v}}&{\small \ion{N}{vi}}&{\small \ion{O}{vii}}  &{\small \ion{Ne}{ix}} &{\small \ion{Mg}{xi}}  &{\small \ion{Si}{xiii}}\\
\hline
{\small {\it resonance} }         & {\small $w$ ($r$)}&{\small 40.279} &{\small 28.792} &{\small 21.603}  &{\small 13.447} &{\small 9.1681} &{\small 6.6471} \\
{\small {\it inter-}}& {\small $x$ }        &{\small 40.711} &{\small 29.074} &{\small 21.796}  &{\small 13.548} &{\small 9.2267}
 &{\small 6.6838} \\
{\small {\it combination}}                                       & {\small $y$}        &{\small 40.714} &{\small 29.076} &{\small 21.799}  &{\small 13.551} &{\small 9.2298} &{\small 6.6869} \\
{\small {\it forbidden}}          &{\small  $z$ ($f$)}   &{\small 41.464}  &{\small 29.531}  &{\small 22.095}  &{\small 13.697} &{\small 9.3134} &{\small 6.7394} \\
\hline
\hline
\end{tabular}
\end{center}
\label{table:lambda}
\end{table}

\subsection{Density diagnostic}\label{sec:density}

\begin{figure}[h]
\begin{tabular}{cc}
\resizebox{4cm}{!}{\includegraphics{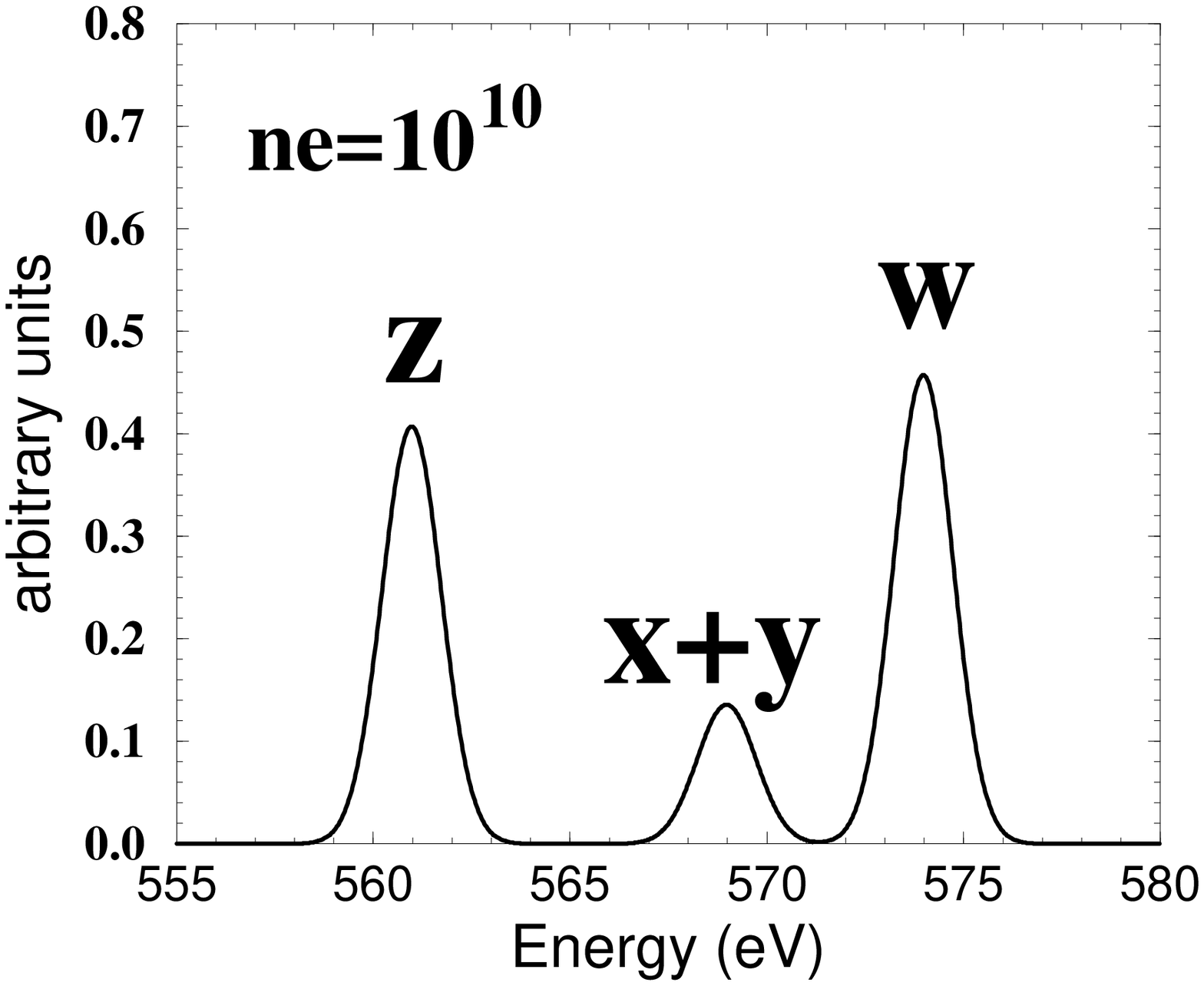}} &\resizebox{4cm}{!}{\includegraphics{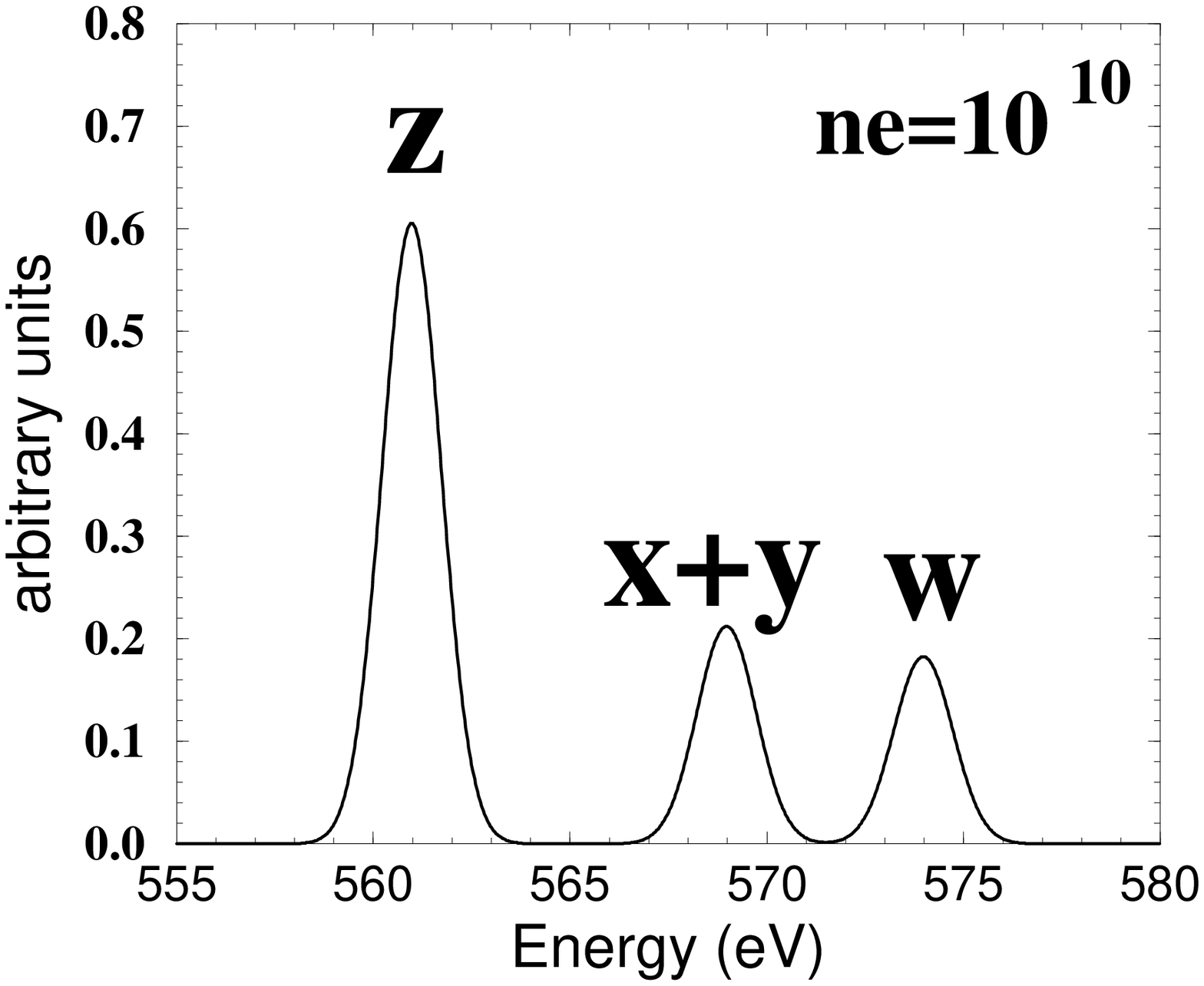}}
\\
\resizebox{4cm}{!}{\includegraphics{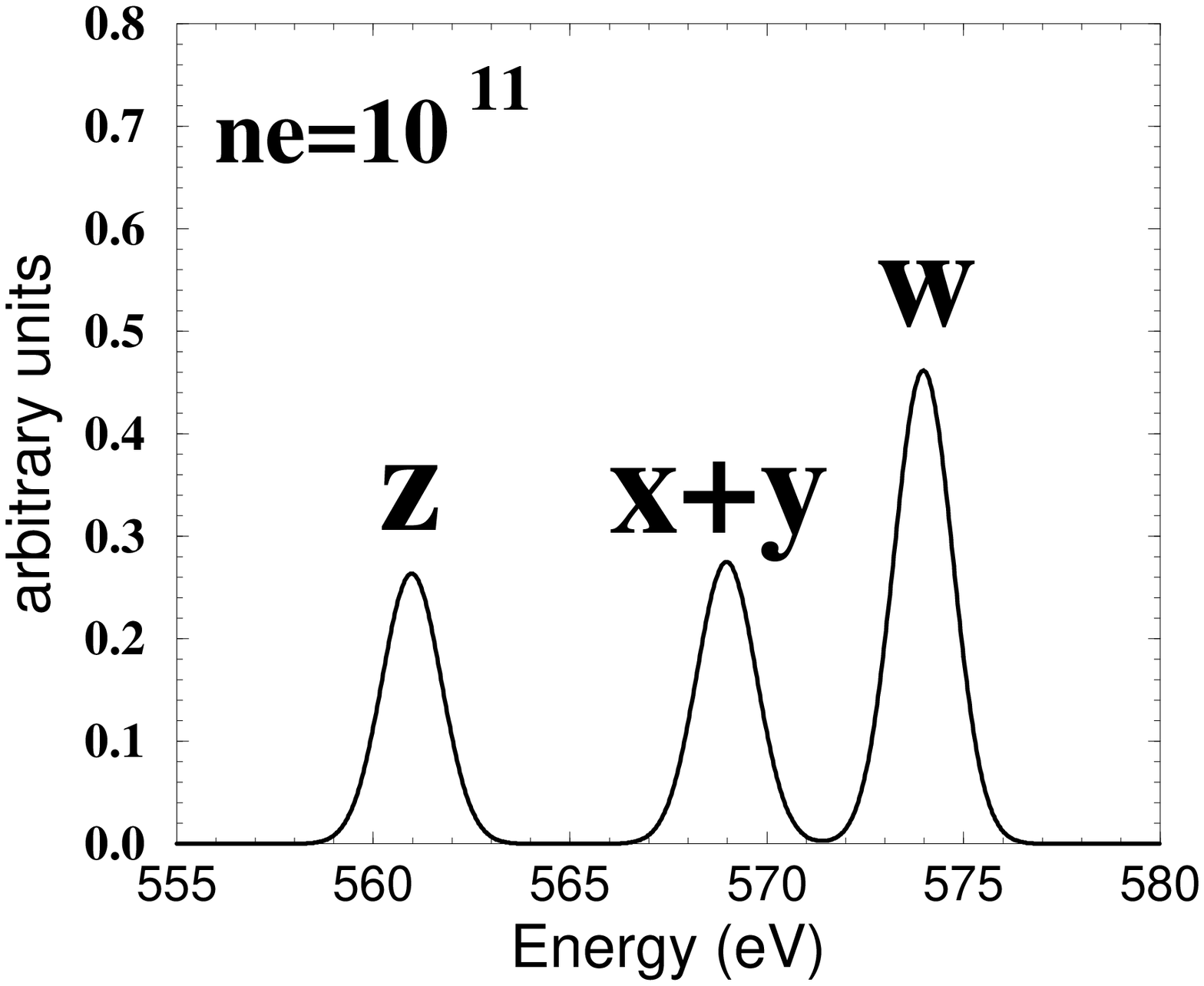}} &\resizebox{4cm}{!}{\includegraphics{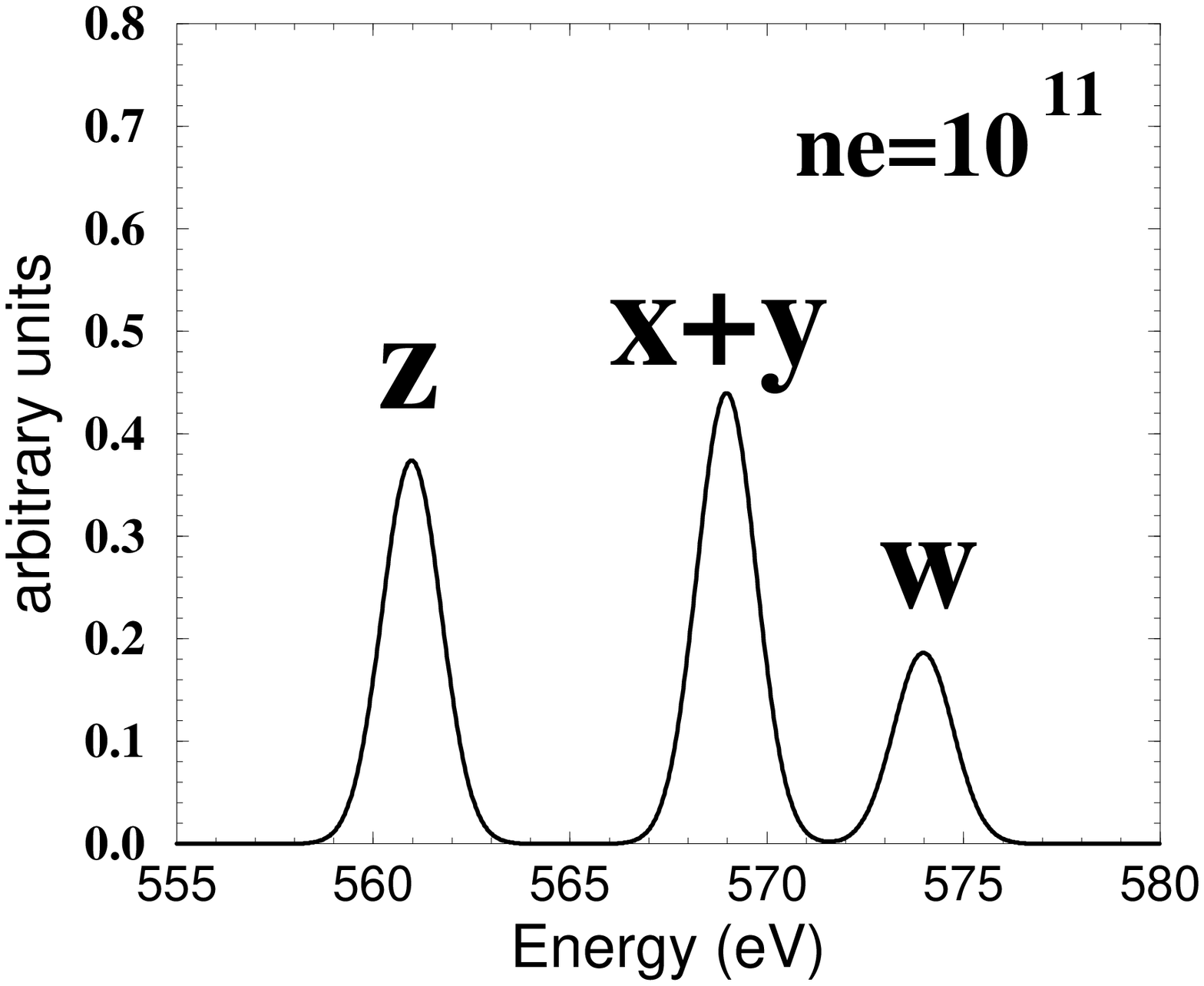}}
\\
\resizebox{4cm}{!}{\includegraphics{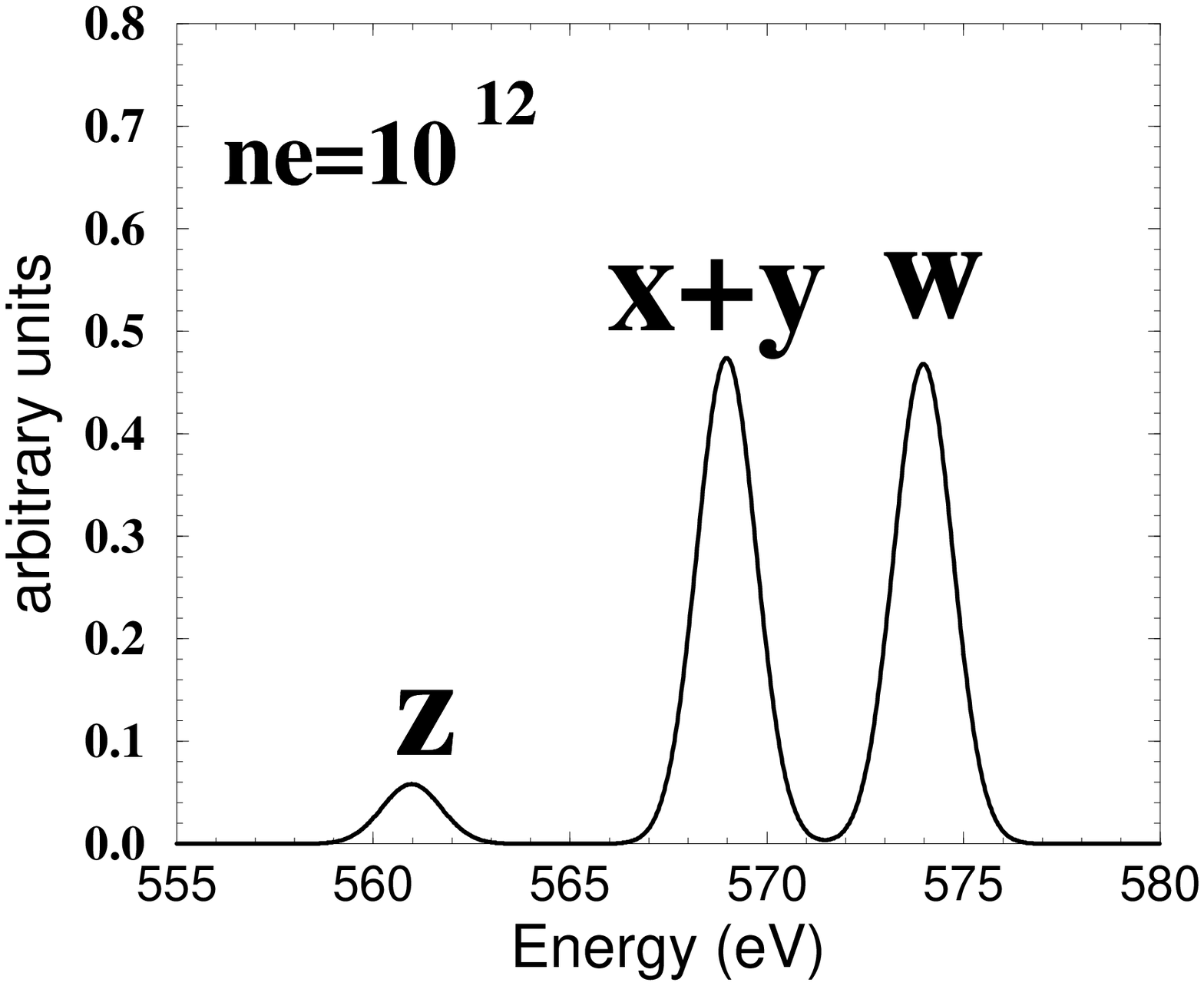}} &\resizebox{4cm}{!}{\includegraphics{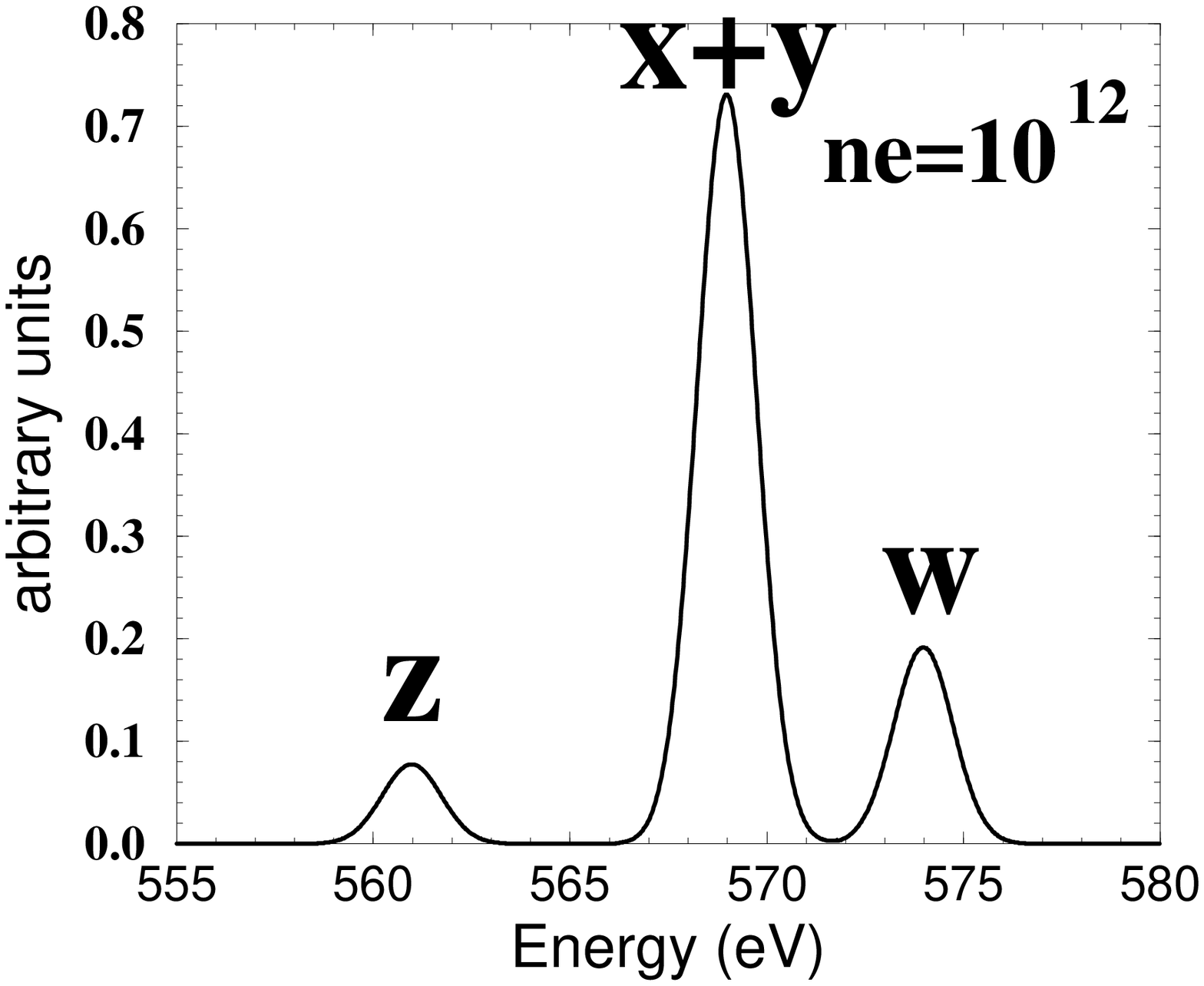}}
\\
\end{tabular}
\caption{\ion{O}{vii} theoretical spectra constructed using the RGS (XMM) resolving power (E/$\Delta$E) for 
three values of density (in cm$^{-3}$). This corresponds (approximatively) to the range where the ratio {\bf R} 
is very sensitive to density. {\bf z}: forbidden lines, {\bf x+y}: intercombination lines and {\bf w}: resonance line. 
{\it At left}: ``hybrid plasma'' at T$_{\mathrm{e}}$=1.5\,10$^{6}$\,K; 
{\it At right}: ``pure'' photoionized plasma at T$_{\mathrm{e}}$=10$^{5}$\,K). 
{\it Note: the intensities are normalized in order to have the sum of 
the lines equal to the unity}.}
\label{fig:XMM}
\end{figure}

\indent In the low-density limit, all $n$=2 states are populated directly 
or via upper-level radiative cascades by electron impact from the He-like 
ground state and/or by (radiative and dielectronic) recombination of H-like 
ions (see Figure~\ref{fig:gotrian}).
These states decay radiatively directly or by cascades to the ground level. 
The relative intensities of the three intense lines are then independent of density. 
As $n_{\mathrm{e}}$ increases from the low-density limit, some of these states 
(1s2s\,$^{3}$S$_{1}$ and $^{1}$S$_{0}$) are depleted by collisions to the nearby 
states where $n_{\mathrm{crit}}$\,C=A, with C being the collisional coefficient rate, 
A being the radiative transition probability from $n$=2 to $n$=1 (ground state), 
and $n_{\mathrm{crit}}$ being the critical density. Collisional excitation depopulates 
first the 1s2s\,\element[][3]{S}$_{1}$ level (upper level of the {\it forbidden} line) 
to the 1s2p\,\element[][3]{P}$_{0,1,2}$ levels (upper levels of the 
{\it intercombination} lines). The intensity of the {\it forbidden} line 
decreases while those of the {\it intercombination} lines increase, hence 
implying a reduction of the ratio $R$ (according to Eq.~\ref{eq:RG}), over approximately 
two or three decades of density (see Fig.~\ref{fig:XMM}).

For much higher densities, 1s2s\,$^{1}$S$_{0}$ is also depopulated to 1s2p$^{1}$P$_{1}$, and the 
{\it resonance} line becomes sensitive to the density.\\
 However caution should be taken for low-Z ions (i.e. \ion{C}{v}, \ion{N}{vi}, \ion{O}{vii}) 
since in case of  an intense UV radiation field, the photo-excitation between the $^{3}$S 
term and the $^{3}$P term is not  negligible. This process has the same effect on the 
{\it forbidden} line and on the {\it intercombination}  line as the collisional 
coupling, i.e. lowering of the ratio $R$, and thus could mimic a high-density plasma. 
 It should be taken into account to avoid any misunderstanding between a high-density 
plasma and a high  radiation field (see $\S$\ref{sec:photo-excitation-dens} for more details).

\subsection{Temperature and ionization process diagnostics}\label{sec:temperature}

\indent The ratio $G$ (see Eq.~\ref{eq:RG}) is sensitive to the electron temperature since 
the collisional excitation rates have not the same dependence with  temperature for the 
{\it resonance} line as for the  {\it forbidden} and {\it intercombination} lines.\\
 In addition, as detailed in Porquet \& Dubau \cite{dporquet-WB2:Porquet00} 
(see also Mewe \cite{dporquet-WB2:Mewe99}, and Liedahl \cite{dporquet-WB2:Liedahl99}), 
the relative intensity of the {\it resonance} $w$ line, compared to the 
{\it forbidden} $z$ and the {\it intercombination } $x+y$ lines, contains  
information about the ionization processes that occur: a strong  {\it resonance} line 
compared to the  {\it forbidden} or the {\it intercombination} lines corresponds 
to collision-dominated plasmas. 
It leads to a ratio of $G=((x+y)+z)/w\sim$1. 
On the contrary, a weak {\it resonance} line corresponds to plasmas dominated  
by the photo-ionization $G=((x+y)+z)/w>$4. An illustration is given Figure~\ref{fig:XMM}.\\
However, as mentionned for the density diagnostic, caution should be taken since 
photo-excitation can mimic a hybrid plasmas, i.e. photo-ionization plus collisional ionization, 
e.g. shock or starburst (see  $\S$\ref{sec:photo-excitation-temp}).

\section{Blended dielectronic satellite lines}
\label{fauthor-E1_sec:fig}

The influence of the blending of dielectronic satellite lines 
for the {\it resonance}, the {\it intercombination} and the 
{\it forbidden} lines has been taken into account 
when their contribution is not negligible in the 
calculation of $R$ and $G$, affecting the inferred electron temperature and density. 
This is the case  for the high-Z ions produced in a collisional plasma, 
i.e. \ion{Ne}{ix}, \ion{Mg}{xi}, and \ion{Si}{xiii} (Z=10, 12, and 14, respectively).

\begin{equation}
R=\frac{z+satz}{(x+y)+satxy}
\end{equation}
\begin{equation}
G=\frac{(z+satz)+((x+y)+satxy)}{(w+satw)},
\end{equation}
\noindent where $satz$, $satxy$ and $satz$ are respectively the contribution 
of blended dielectronic satellite lines to the 
{\it forbidden} line, to the {\it intercombination} lines, and to the {\it resonance} line, respectively. 
One can note that at very high density the $^{3}$P levels are depleted to the $^{1}$P level, and in that 
case {\it x+y} decreases and $R$ tends to $satz$/$satxy$.\\

The intensity of a dielectronic satellite line arising from a doubly excited state with principal quantum number $n$ 
in a Lithium-like ion produced by dielectronic recombination of a He-like ion  is given by: 
\begin{equation}
I_{s}=N_{\rm He}~ n_{\mathrm e}~ C_{s},
\end{equation}
where $N_{\rm He}$ is the population density of the considered He-like ion in the ground state 1s$^{2}$ with statistical
weight $g_1$ (for He-like ions $g_1=1$).\\
 The rate coefficient (in cm$^{3}$\,s$^{-1}$) for dielectronic recombination is given by (Bely-Dubau et al. 
 \cite{dporquet-WB2:Bely79}):
\begin{equation}
C_{s}=2.0706\ 10^{-16}~\frac{e^{-E_{s}/kT_{\mathrm e}}}{g_{1} T_{\mathrm e}^{3/2}}~F_{2}(s),
\end{equation}
where $E_{s}$ is the upper energy level of the satellite line $s$ with statistical weight $g_s$ above the  
ground state 1s$^{2}$ of the He-like ion.
 $T_{\mathrm e}$ is the electron temperature in K, and 
$F_{2}(s)$ is the so-called line strength factor 
(often of the order of about 10$^{13}$~s$^{-1}$ for the stronger lines) given by
\begin{equation}
F_{2}(s) = {{g_s A_a A_r} \over {(A_a + \sum A_r)}},
\end{equation}
where $A_a$ and
$A_r$ are transition probabilities (s$^{-1}$) by auto-ionization and radiation,
and the summation is over all possible radiative transitions from the satellite level $s$. \\

 Since the contribution of the blended dielectronic satellite lines depends on the spectral resolution
 considered, we have estimated the ratios $R$ and $G$ for four specific spectral resolutions (FWHM): 
RGS-1 at the first order, LETGS, HETGS-MEG, and HETGS-HEG (Porquet, Mewe et al \cite{dporquet-WB2:Porquet01}).
At the temperature at which the ion fraction is maximum for the He-like ion 
(see e.g. Mazzotta et al. \cite{dporquet-WB2:Mazzotta98}), the differences between the 
calculations for $R$ (for $G$) with or without taking into account the blended dielectronic satellite lines are only
of about 1$\%$ (9$\%$), $2\%$ (5$\%$), and 5$\%$ (3$\%$) for \ion{Ne}{ix}, \ion{Mg}{xi}, and 
\ion{Si}{xiii} at the low-density limit and for $T_{\mathrm rad}$=0\,K, respectively. 

However, for photo-ionized plasmas where recombination prevails and the temperature 
is much lower (e.g., T$\la$0.1T$_{m}$), the effect on $R$ and $G$ can be much bigger 
since I$_{sat}$/I$_{w}$$\propto T^{-1}{\rm e}^{{(E_{w}-E_{sat}})/k{\rm T}}$.
For very high density $n_{\mathrm e}$ the contribution of the blended dielectronic 
satellite lines to the forbidden line leads to a ratio $R$ which tends to $satz/satxy$, 
hence decreases much slower with $n_{\mathrm e}$ than in the case where the contribution 
of the blended DR satellites is not taken into account.

\newpage
\section{Optical depth}

If the optical depth of the resonance line is not taken into account, the calculated ratio G could be 
overestimated (inferred temperature underestimated) when the optically-thin approximation 
is no longer valid. This has been estimated with an {\it escape-factor method}, 
e.g., for the case of a {\it Warm Absorber in an AGNs} 
(Porquet, Kaastra, Mewe, Dubau \cite{dporquet-WB2:Porquet02}).

\section{Radiation field (photo-excitation)}\label{sec:photo-excitation}

\subsection{Influence on density diagnostic}\label{sec:photo-excitation-dens}

\noindent A strong radiation field can mimic a high density if the photo-excitation 
$^{3}$S$_1$ level ({\it f} line) $\to$ $^{3}$P$_{0,1,2}$ levels ({\it i} lines) exceeds 
the electron collisional excitation. ex: $\zeta$ Puppis (Kahn et al. \cite{dporquet-WB2:Kahn01}, 
Cassinelli et al. \cite{dporquet-WB2:Cassinelli01}). Rate of photo-excitation (in s$^{-1}$) 
(Mewe \& Schrijver \cite{dporquet-WB2:Mewe78a}) in a stellar photospheric radiation field 
with effective black-body radiation temperature ${T_{\mathrm rad}}$ is written
 as: 
\begin{equation}
B_{mp_k} = {\frac{W A_{p_km} (w_{p_k}/w_m)}{exp \Bigl(\frac{\Delta E_{mp_k}}{kT_{\mathrm rad}}\Bigr) - 1}},
\end{equation}
where $A$ and $B$ are the Einstein coefficients and the radiation is diluted by a factor $W$ given by
\begin{equation}
W=\frac{1}{2}~\left[1-\left(1-\left(\frac{r_{*}}{r}\right)\right)^{1/2}\right],
\end{equation}

\noindent $\bullet$ {\bf W=1/2} (close to the stellar surface, $r=r_*$; e.g., \object{Capella} 
and \object{Procyon}: Audard et al. \cite{dporquet-WB2:Audard01}, 
Mewe et al. \cite{dporquet-WB2:Mewe01}, Ness et al. \cite{dporquet-WB2:Ness01}).\\
$\bullet$ {\bf W$<<$ 1/2} (radiation originates from another star at larger distance; 
e.g., Algol, where K-star is irradiated by B-star, $W \simeq 0.01$: Ness et al. \cite{dporquet-WB2:Ness02}).\\ 

\begin{figure}
\hspace*{1cm}\includegraphics[width=6cm,angle=-90]{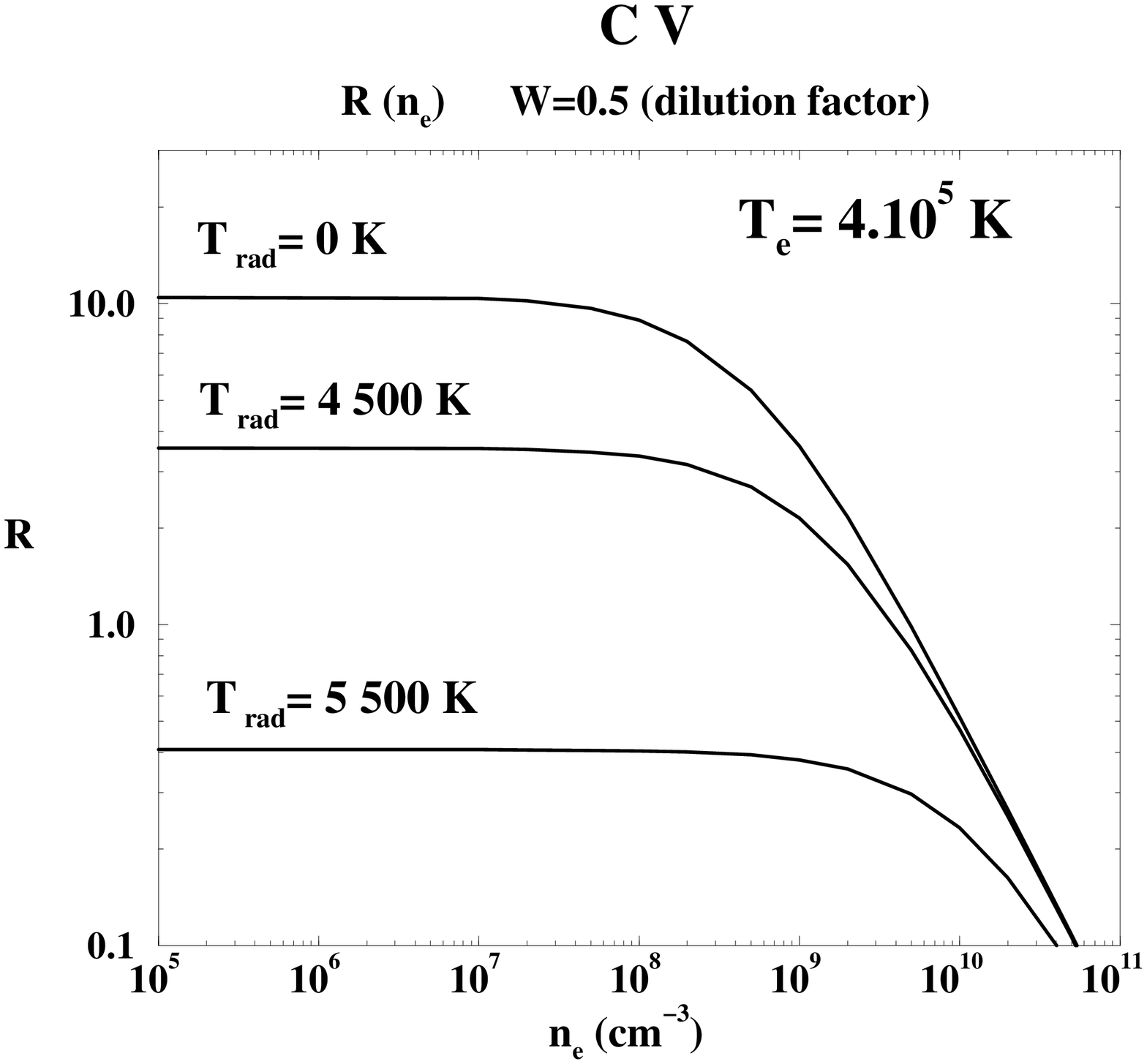} \\
\hspace*{1cm}\includegraphics[width=6cm,angle=-90]{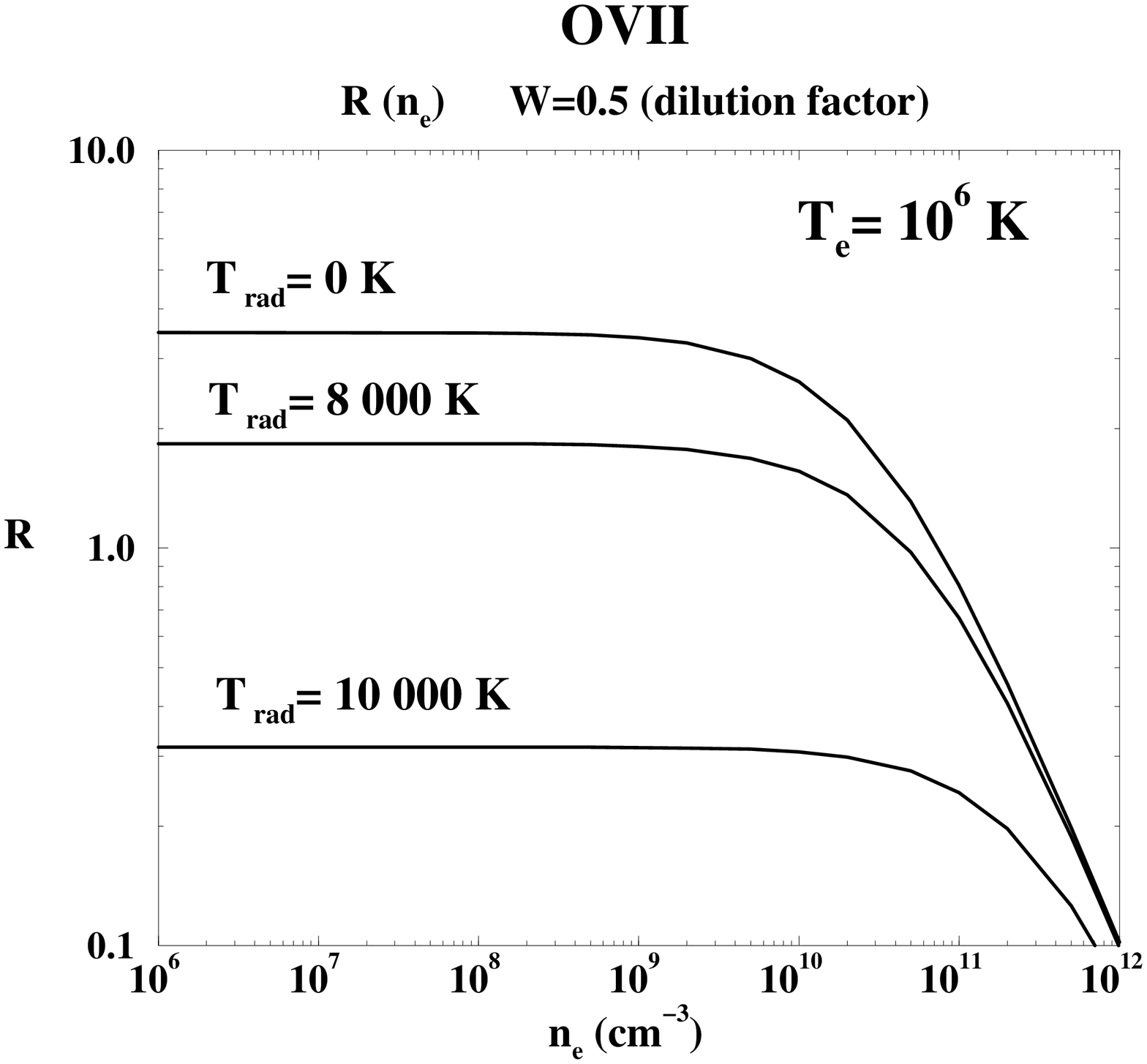} \\
\caption{ Ratio $R(n_{e})$=$z$/$(x+y)$ in dependence 
of stellar radiation field for \ion{C}{v} (top) 
and \ion{O}{vii} (bottom). 
Note that, e.g. for \ion{O}{vii} a radiation 
field with effective temperature of 10$^{4}$\,K 
can mimic a density of about 3\,10$^{11}$\,cm$^{-3}$.}
\label{fig:fig1}
\end{figure}

Porquet et al. (2001) showed that photo-excitation is important for \ion{C}{v}, \ion{N}{vi}, 
\ion{O}{vii} for T$_{\mathrm rad}\geq$(5-10) 10$^3$\,K  (see Fig.~\ref{fig:fig1}), 
and for higher-Z ions when T$_{\mathrm rad}\geq$ few 10$^4$~K.\\

\subsection{Influence on ionization process diagnostic}\label{sec:photo-excitation-temp}

Recently,  Kinkhabwala et al.  (2002),  pointed out the important effect of
the photo-excitation in the high Rydberg series lines.  Indeed, the ratio of high-$n$ 
lines to Ly$\alpha$ bring evidence for photo-excitation in Warm Absorber Seyfert\,2. 
They clearly showed that in addition to photo-ionization 
(treated in Porquet \& Dubau \cite{dporquet-WB2:Porquet00}), 
 photo-excitation process is sufficient to fit the data of Seyfert  galaxies 
without needing an additional collisional ionization process (e.g. shock or starburst).
 Then both photo-ionization and photo-excitation are needed to inferred 
unambiguously the ionization process occurring in the plasmas.
\newpage
\section{Conclusion}

 Helium-like density and temperature diagnostics has now become 
a powerful tool in the analysis of the high resolution {\sl Chandra} 
and {\sl XMM-Newton} X-ray spectra (Porquet \& Dubau 2000). 
Therefore, we have revisited the calculations of the  ratios 
$R=z/(x+y)$ and $G=((x+y)+z)/w$ of the 
{\it $z$, $(x+y)$, and $w$ ``triplet'' lines of the He-like ions} 
\ion{C}{v}, \ion{N}{vi}, \ion{O}{vii}, 
\ion{Ne}{ix}, \ion{Mg}{xi}, \ion{Si}{xiii},
 taking into account all relevant processes and improved atomic data.\\
The calculations were done for optically thin 
plasmas in collisional ionization 
equilibrium (e.g., stellar coronae: 
Porquet, Mewe et al. \cite{dporquet-WB2:Porquet01}).
The influence of an external radiation field on the depopulation of the upper 
level of $z$ is considered which can be important for hot OB or F stars 
(e.g., \object{$\zeta$ Puppis}, \object{Procyon}, and \object{Algol}).
In preparation are improved calculations for photo-ionized and hybrid 
plasmas (e.g., warm absorber in AGNs: Porquet, Kaastra, Mewe, Dubau \cite{dporquet-WB2:Porquet02}), 
and will be extended to transient ionization plasmas 
(young SNRs: Kaastra, Mewe, Porquet, Raassen \cite{dporquet-WB2:Kaastra02}), 
where inner-shell ionization of the Li-like ion can contribute 
significantly to the intensity of the forbidden line 
(see also Mewe  \cite{dporquet-WB2:Mewe02}).\\

%\newpage
\begin{acknowledgements}

D.P. acknowledges grant support from the ``Institut National des Sciences de l'Univers'' and from the
``Centre National \linebreak d'Etudes Spatial'' (France). 
The Space Research Organization Netherlands (SRON) is supported financially by NWO.\\
\end{acknowledgements}


\begin{thebibliography}{}


\bibitem[2001]{dporquet-WB2:Audard01}
Audard, M. et al. (2001), A\&A, 365, L329
\bibitem[1979]{dporquet-WB2:Bely79}
Bely-Dubau, F., Gabriel, A. H., Volont{\'e}, S. 1979, MNRAS, 189, 801
\bibitem[2001]{dporquet-WB2:Cassinelli01}
 Cassinelli J. P., Miller N. A., Waldron W. L., MacFarlane J. J., Cohen D. H. (2001), ApJ, 554, L55
\bibitem[1980]{dporquet-WB2:Doyle80}
Doyle J.G. (1980), A\&A, 87, 183
\bibitem[1969]{dporquet-WB2:Gabriel69}
Gabriel A.H. \& Jordan, C. (1969), MNRAS, 145, 241
\bibitem[2002]{dporquet-WB2:Kaastra02}
Kaastra J. S, Mewe R., Porquet D., Raassen A. J. J. (2002), in preparation (Paper IV)
\bibitem[2001]{dporquet-WB2:Kahn01}
Kahn S.~M. et al. (2001), A\&A, 365, L312
\bibitem[2002]{dporquet-WB2:Kinkhabwala02}
Kinkhabwala et al. 2002, these proceedings
\bibitem[1999]{dporquet-WB2:Liedahl99}
Liedahl D.~A., 1999, X-Ray Spectroscopy in Astrophysics,  189
\bibitem[1998]{dporquet-WB2:Mazzotta98}
 Mazzotta P. et al. 1998, A\&AS, 133, 403
\bibitem[1978a]{dporquet-WB2:Mewe78a}
Mewe R. \& Schrijver J. (1978a), A\&A, 65, 99
\bibitem[1978b]{dporquet-WB2:Mewe78b}
Mewe R. \& Schrijver J. (1978b), A\&A, 65, 115 
\bibitem[1978c]{dporquet-WB2:Mewe78c}
 Mewe R. \& Schrijver J. (1978c), A\&AS, 45, 11 
\bibitem[1999]{dporquet-WB2:Mewe99}
Mewe R., 1999, X-Ray Spectroscopy in Astrophysics,  109
\bibitem[2001]{dporquet-WB2:Mewe01}
 Mewe R., Raassen A.J.J., Drake J.J., Kaastra J.S., van der Meer R.L.J., Porquet D. (2001), A\&A, 368, 888 
\bibitem[2002]{dporquet-WB2:Mewe02}
Mewe R. 2002, these proceedings
\bibitem[2001]{dporquet-WB2:Ness01}
Ness J.-U., Mewe R., Schmitt J.H.M.M., Raassen A.J.J., Porquet D. et al. (2001), A\&A, 367, 282
\bibitem[2002]{dporquet-WB2:Ness02}
Ness J.-U. et al. (2002), in preparation
\bibitem[2000]{dporquet-WB2:Porquet00}
Porquet D. \& Dubau J. (2000), A\&AS, 143, 495 (Paper I)
\bibitem[2001]{dporquet-WB2:Porquet01}
Porquet D., Mewe R., Dubau J., Raassen A. J. J., Kaastra J. S (2001), A\&A, 376, 1113 (Paper II)
\bibitem[2002]{dporquet-WB2:Porquet02}
 Porquet D., Kaastra J. S, Mewe R., Dubau J. (2002), in preparation (Paper III)
\bibitem[1981]{dporquet-WB2:Pradhan81}
Pradhan A. K. \& Shull J. M. (1981), ApJ, 249, 821
\bibitem[1978]{dporquet-WB2:Vainshtein78}
Vainshtein L.~A., Safronova U.~I., 1978, Atomic Data and Nuclear Data Tables,  21, 49
\end{thebibliography}
\end{document}